# Key Management in Mobile Sensor Networks

Anvesh Reddy Aileni

Computer Science Department

Oklahoma State University, Stillwater, OK - 74078

anvesh@okstate.edu

**ABSTRACT**

Wireless sensor networks consist of sensor nodes with limited computational and communication capabilities. This paper deals with mobile sensors which are divided into clusters based on their physical locations. Efficient ways of key distribution among the sensors and inter and intra cluster communications are examined. The security of the entire network is considered through efficient key management by taking into consideration the network's power capabilities.

1. **INTRODUCTION**

A sensor network consists of spatially distributed nodes with limited computational and communication capabilities. These are used in varied applications such as monitoring environmental conditions, military and other industrial purposes. The nodes which are distributed spatially have to communicate with each other in a secured manner since the data associated with these nodes may be confidential. Efficient cryptography techniques are to be used for communication among the nodes. Use of public key cryptography is omitted here keeping in mind the computational limitations of sensor nodes [1]. Also, using a single key for the entire network will compromise on the security of the system as a threat on even one node would bring the system down.

We assume that each sensor runs on battery power and is equipped with a certain kind of radio communication device which can only communicate in a small range. We also assume devices that use dual radio communication, one for short range and other for long range. These kinds of dual radio communication devices are produced by Sensoria Corporation [14]. These devices consume less computational power and the operating system which is used is less complex that the normal one. Examples of operating systems that can be used is Tiny OS[15] and Lite OS[16]. Algorithms used for computational purpose should also have to consume less energy and therefore use of certain cryptographic methods is ruled out.

Due to the storage limitations of the sensors nodes, all unique keys cannot be stored. Consider a total of *m* nodes out of which two nodes share a unique key. This will result in a total of $\left(\frac{m(m-1)}{2}\right)$ keys in the network with each node storing $(m-1)$ keys. This is very high owing to the storage constraints of a node if *m* is large. Also, considering the ad-hoc nature of the sensor networks, some nodes will continue to be added to the network over time. Thus efficient key management has to be performed in order to store keys in nodes such that the network security is not compromised and also the total number of keys stored in the node is reduced.

Key management becomes tougher when the network we deal contains sensors which are mobile. These sensors tend to move in any direction making the key distribution more complex. So the control station -- here we call it *base station* -- which actually monitor the network has to have the idea of where each sensor is and with what other sensors it has to share keys in order to have secure communication. Most key management schemes either deal with networks which have completely static sensors or networks which have partial static and partial mobile. But in the scheme considered in this paper, all the sensors are capable of roaming in any direction.

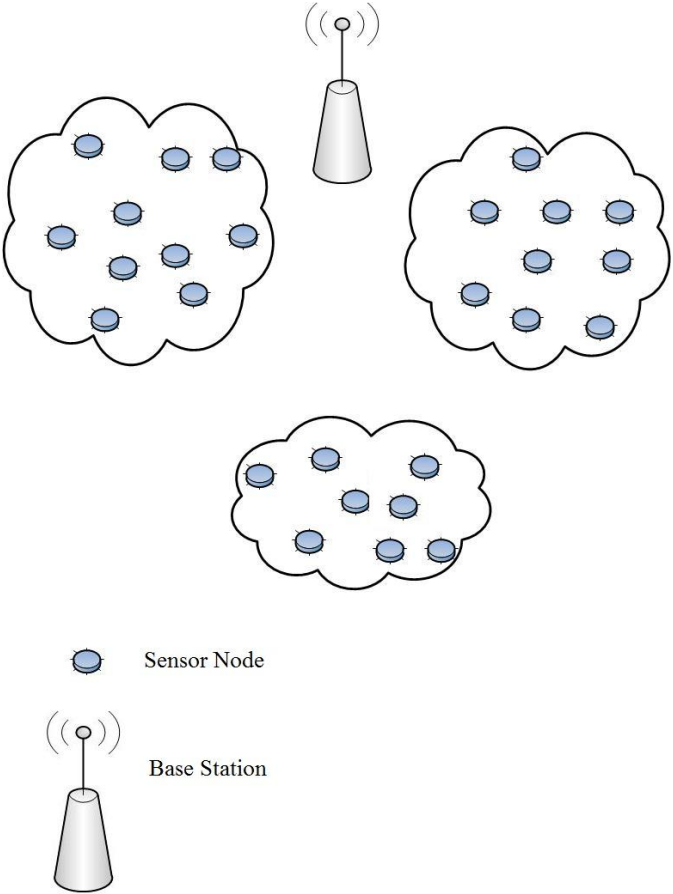

Figure 1 Basic idea of clustered network

The key ring approach requires that data be routed over many nodes before it reaches its destination. This indicates that there is an underlying connection between routing and security and that implicit security technique [3],[4],[5] may be used to route data via multiple channels, thus spreading the vulnerability over several nodes and communication channels making an adversary's task harder. There is also the question of node compromise which may be countered by using tamper-resistance hardware in each node [6]. Further, overlay architecture has been used to achieve balance between number of keys per node and routing complexity [13]. Other methods of potential applications to mobile node security include [6], [22],[23].

## 2. LITERATURE REVIEW

Eschenauer and Gligor [2] proposed a "key ring" scheme using random graphs wherein from a large pool of keys *S* few keys were selected and stored in the node such that two neighbors share at least a key with certain probability *p*. Each node carries a subset of keys called the key ring of size *m* from *S* and finds common set of keys among subset of any two randomly chosen key rings. This scheme requires three main steps that of key pre-distribution, key discovery and path key identification. In the key discovery phase every node sends a message to all other nodes. The neighbors upon receiving the message determine which key are shared with which node. Also a path is established among nodes if they do not share a key and this is done in path key identification. This populates the entire network and this process may have to be repeated every time a node is added or a node is replaced. Q-composite proposed by Chen *et. al* [17] improves the scheme by extending the idea of EG by overcoming the difficulties when two sensors do not share a common key. These two concepts are not practical for use in mobile sensors.

Dongg et al. [18] proposed group based key pre-distribution scheme. This scheme has two parts: one is group based EG in which key pool size at any instance of time is 500 and each sensor node randomly selects 50 keys each from in-group and cross-group instance., whereas the second one is a group based PB scheme that uses a 49 degree bivariate polynomial. Polynomial share from in-group and cross-group is assigned to each sensor node. Lin et al. [21] have proposed a scheme which divides the entire network in clusters and each cluster has a cluster head and sensors and a base station for the network. This scheme use LEACH protocol in electing new cluster heads. Cluster heads are ordinary sensor nodes which could communicate to base stations and all the cluster heads in single hops.

Location based protocol proposed by Kihyun et al.[19] uses three phases in establishing efficient routing in mobile sensor networks. Although this scheme does not mention key management, it has a method for establishing routing protocol among mobile sensors. In the first phase, the node sends advertisements message to all the neighbor nodes and then receives reply from them in the second phase if only when they want to send the reply. In the third phase, data is forwarded upon receiving the reply message from sensors. This way communication is made and messages are exchanged among nodes. This uses global positioning system through which each sensor knows about the location of itself and also the location of sink node. Kifayat et al. [20] proposed group based key management for mobile sensor networks in which they have both static and mobile sensor nodes and the network is divided into groups. Key ring k, from pool of key S, are loaded into static and mobile sensors in pre-deployment phase in a similar fashion as employed in [2]. With these keys they can communicate with sensors in an out of each group and moreover each group has a group leader that receives data in encrypted format from nodes in the group and sends to sink node. Mobile nodes roam in the network and establish connection with at least one sensor in each group with a certain probability within its communication range. This scheme has one major drawback as when group leader compromise entire group of sensor nodes get compromised.

## 3. PROPOSED SCHEME

Our proposed network has been designed considering a battlefield scenario. Since soldiers or military vehicles move in groups we have designed the network based on groups. Thus the entire network of sensors is divided into clusters based on their location and also based on the neighboring nodes. Each cluster has a cluster head that is an ordinary node. In designing this network, we initially considered sensors having two kinds of radio communications, one is short range and other is long range. These kinds of sensors are being produced and are available for use by Sensoria Corporation [14].

We use *K-means++* clustering algorithm in clustering the network and we assign a sensor node as the cluster head which is at the mean position identified. Sensors when chosen as cluster heads switch on their long range radio communication to be able to communicate with base station and other cluster heads. Thus a cluster is formed with certain nodes assigned by the *k-means++* algorithm and they tend to move in and out of cluster frequently. Clustering is done again in order to identify new clusters and cluster heads.

After clustering, next part is to distribute keys among the sensors. Key pre distribution can be eliminated from here as, at any moment of time, each sensor might not have the same neighbors. So the keys, pre-stored before deployment, may not have any keys useful to be shared with sensors in the communication range. The keys have to be distributed dynamically, that is, they have to be sent by the base station each time it changes it position and completely moves out of a cluster. For this, each sensor stores a key shared with the base station which is used to exchange the keys in a plaintext format. Each time a sensor joins a new cluster, its new key set is transferred to the sensors wiping out all the old keys. In addition to these, a cluster head also stores keys of other cluster heads in its communication range. Birthday paradox has been employed here in order to store keys in each sensor.

### 3.1. CLUSTERING

In *k*-means++ clustering algorithm, the number of clusters is fixed *a priori*. We assume that the number of clusters to be chosen is *k* and this choice is based on the network size and geographical positions of each node. Consequently, we will have *k* sub controllers monitoring nodes in their respective clusters.

According to *k*-means++ clustering algorithm, initially *k* means are to be identified, one for each cluster. For this, as explained in *k*-means++ approach, following steps are to be followed,

1. Choose an initial centre *x*, from the set of all node points say μ, uniformly at random.
2. Choose next mean from the set and name it *y* with probability $\frac{D(y)^2}{\sum_{x \in \mu} D(x)^2}$ proportional to $D(x)^2$
3. Repeat step 2 until all the *k* centers are identified.

Once these centers are identified, these would be the means of the clusters and each node distance from all the means is calculated and is associated to the nearest mean in a process called binding. This process is repeated up to the stage where all the positions of nodes in the network are visited at least once. After this, $k$ new means are calculated. Once the positions of $k$ new means are identified, binding is done again. This process is repeated until there is no change in the location of means. The main goal is to minimize the sum of squares of distance within each cluster given by (1)

$$\sum_{i=1}^{k} \sum_{x_j \in S_i} ||x_j - m_j||^2 \longrightarrow (1)$$

Where $m_1, m_2, m_3 \ldots m_k$ are the means of the clusters initially defined.

$$S_i^t = \{ x_j : || x_j - m_i^t || \leq || x_j - m_{i*}^t || \text{ for all } i* = 1, 2, \ldots k \}$$

$x_j$ is the relative distance of $(x,y)$ in the network from origin.

$m_i$ is the relative distance of mean (x,y) from origin.

Updating the means:

$$m_i^{t+1} = \frac{1}{|S_i^t|} \sum_{x_j \in S_i^t} x_j$$

$t$ – Ranges from 1, 2... till the means does not change

$j$ – 1, 2 ..... n.

$i$ – 1, 2 ..... n.

Thus, the $k$ means locations will be identified and these will remain as locations for the cluster heads.

### 3.2 KEY DISTRIBUTION

In this scheme, the birthday problem is used in distributing keys among nodes in the network [8],[9],[10]. According to the birthday problem we have to find number of people with certain probability, so that at least two among them share a common birthday.

Let the probability with which we have to find number of people $n$ in a room to be $p(n)$ and number of days to be $d$. So according to the birthday problem

$$p(n) = 1 - \frac{n! \binom{d}{n}}{d^n}$$

The equation can be approximated to

$$p(n) = 1 - \left(1 - \frac{1}{d}\right)^{C(n,2)}$$

We can draw an analogy to the present scenario as follows,

1) The number of nodes in a cluster at any moment of time with total number of days.
2) The number of people in a room can be associated to number of other sensors with which each sensor at a particular moment of time shares common unique key.

Assume total number of nodes to be *m* in the network and the network is divided into *k* clusters. Choose randomly a cluster with *n* nodes with probability *p(s)*, given by

$$p(s) = 1 - \left(1 - \frac{1}{n}\right)^{C(s,2)}$$

Here, *s* is the number of nodes with which any node shares a key.

The key set of each sensor is refreshed to new set of keys based on the location of neighboring sensors. These new set of keys is transferred by the base station based on the sensor's current location. For the base station to securely transfer set of keys to the sensor, each sensor is preloaded with a key which is used in secure communication with the base node.

### 3.3 COMMUNICATION

Communication is of two types, one is among sensors within the clusters and among sensors belonging to different clusters. For sensors within the cluster, it is the normal exchange of messages as it is assumed that when a cluster is formed the maximum distance between any two points is the communication range of a sensor. That is, a sensor has a potential of sending message to any other sensor in a single hop provided it shares a common key. For communication among nodes belonging to different clusters, the source sensor sends message to the cluster head initially. The cluster head based on the destination sensor's ID identifies to which cluster it belongs to and then sends message to that cluster head. It may be a single hop communication or a multi hop communication between two cluster heads. Once the destination cluster head receives the message, it forwards the message to the destination sensor.

Assume a node, say *i*, wants to send message to any node say *j* which is within the cluster, then using following procedure it would send the message.

{n | if n ∈ {key set of i} and  n where $d_{nj} < d_{xj}$, x = {key set of i}}

*$d_{nj}$* - define a node *n* which is closest node possible to *j*.

It means node *i* sends message to node *n* which close to *j*, if it does not share common key with *j*. Then node *n* checks for other node and this process is repeated till a node is found which exchange key with *j*. This procedure is used in communication among source and destination sensors. This is possible since all the nodes within a cluster can communicate with all other nodes. A similar method is employed for communication among cluster heads.

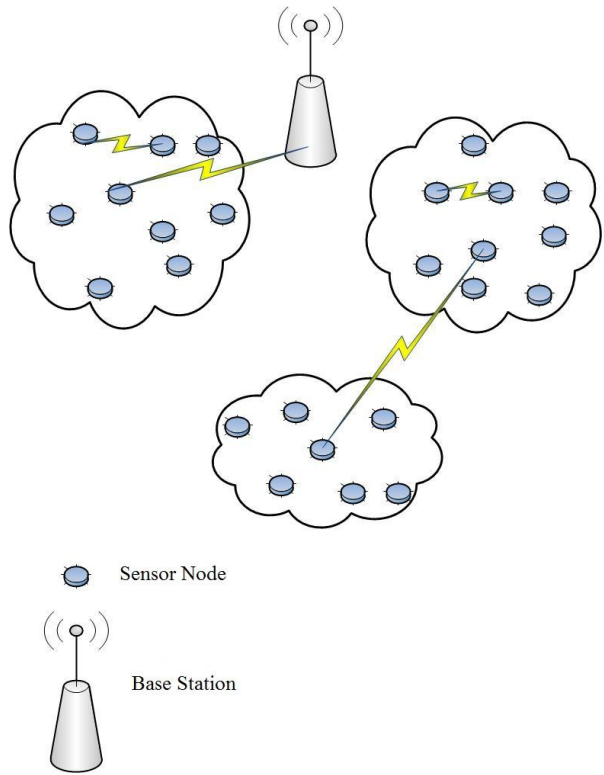

Figure 2 Figure showing two kinds of communications

## 4. ADDITION AND DELETION OF NODES

When the nodes are added, the information is first updated to the base station and based on its location it joins any cluster. If new node is not within the range of any cluster, then that node will form a new cluster with zero other nodes in it. This node will switch on its long range radio communication for communicating with other nodes. As these nodes are mobile, this node at any moment of time may join in any of the cluster and then it can turn off its long range radio thus reducing the power overhead. When the nodes, either loses their battery power or compromised or is completely dead by any reason, we assume that the node has been deleted. The deleted node information is updated to all the nodes within its cluster. A new set of keys are generated and normal operation is resumed. If it is the head node, then new head node is selected using clustering algorithm and this information is sent to all nodes by the base station.

## 5. SIMULATION AND RESULTS

Simulations were carried out for a network of 12000 nodes with a maximum of 300 and minimum of 200 nodes in each cluster. At different percentages, simulations were carried out and each time simulation was run for 30 times. As discussed, each sensor has potential of communicating with any other sensor in a cluster in single hop. According to birthday problem, the number of keys each sensor can store at 20 percent is 8 and at 90 percent it is 37. Addition to these, if a sensor is cluster head it has to store an additional of 3 to 10 keys to be able to communicate with other cluster heads. Figure 1 is graph showing the number if hops at different percentages. It can be observed clearly that upon increase in the number of key stored, the number of hops will decrease.

By clustering we are actually limiting the maximum key pool size i.e., the number of nodes which can share common keys are only among the sensor within a cluster but not among the sensors in the network. When the sensors moves out of the cluster and joins any other cluster, the new key set will be transferred deleting the previous set of keys. Thus by this the memory over head on each sensor has been reduced.

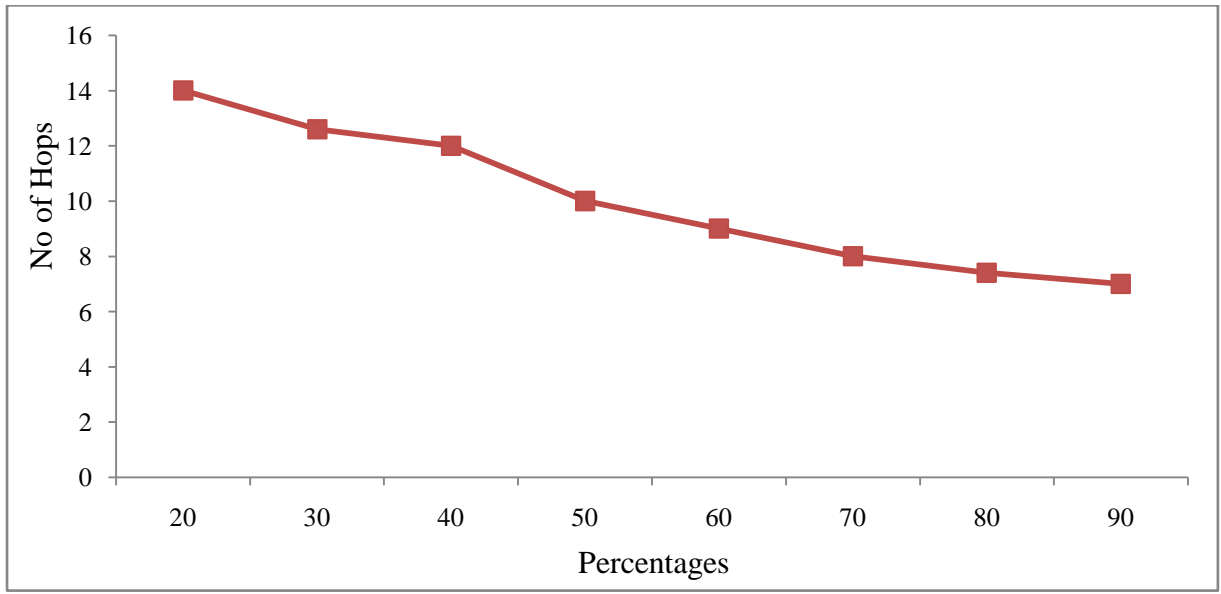

Figure 3 No of Hops vs Percentage

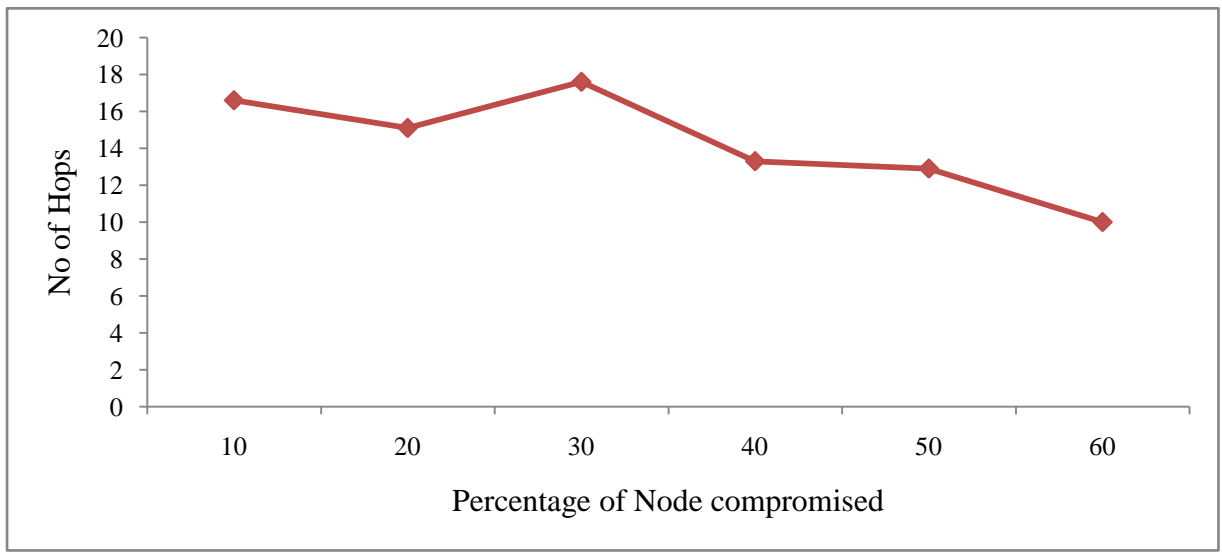

Figure 4 No of Hops vs Percentage of nodes compromised

Simulations were also carried on to check the resilient nature of the network. This was done by removing a few sensors from the cluster and then carrying on the simulation. Figure 4 is a graph representing the number of hops to the percentage of nodes compromised. It can be observed comparing figure 3 and 4 that the number of hops overall have been increased but fall in the number of hops upon increase in number of compromised nodes. This is because of decrease in total number of sensors.

Figure 5 shows us the number of unreachable nodes at different percentage of compromised nodes. If 20 percent of overall nodes in a cluster get compromised, then 2 percent of nodes are unreachable. Similarly when 70 percent of nodes are compromised, then 5.6 percent of nodes are unreachable. And if more than 70% of the nodes in the network are compromised, then the network is assumed to be compromised.

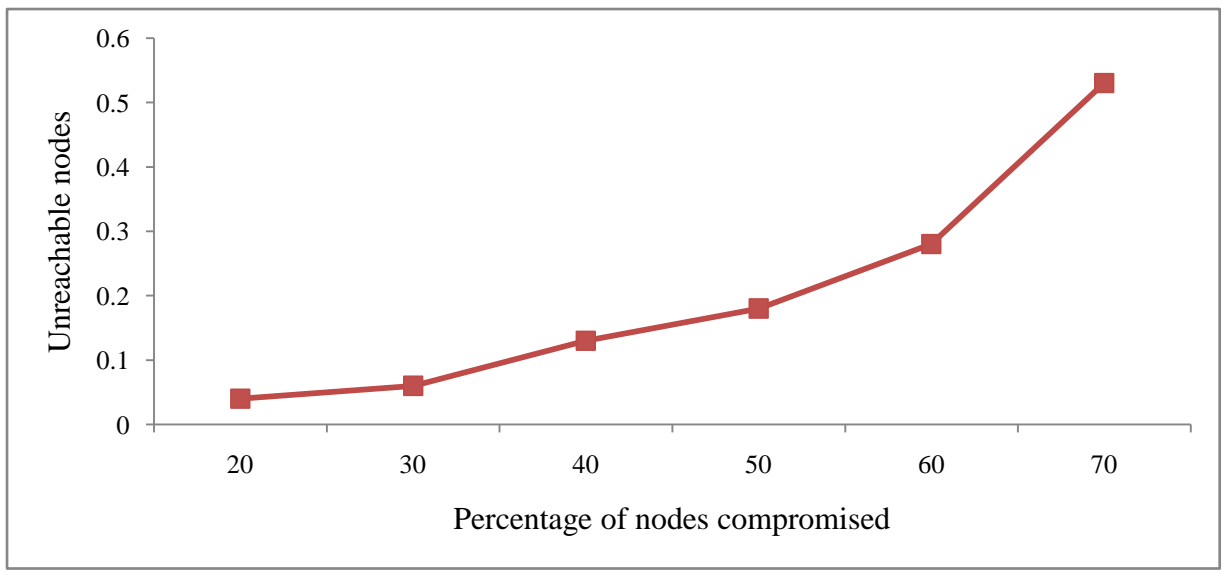

Figure 5 No of unreachable nodes vs Percentage of nodes compromised

Most of the time, a sensor operates only with the short range radio and therefore overall power dissipation is not high. But when the sensor is a cluster head, it operates on its long range radio thus making it to consume more power. If a particular sensor remains operating on its long range radio then there is a chance that it loses its battery power soon. Therefore, cluster head has to rotate in order to distribute the overall power consumption of the network. Since sensors in the network are mobile they tend to change their position very often and so is change of role of cluster head to all the sensors in the cluster.

## 6. **CONCLUSION**

The scheme presented in this paper is an effective ways of distributing keys among mobile sensors. The network is divided into clusters based on their physical location and the maximum distance between any two points in the cluster is the communication range of the sensor. By dividing the network into clusters we are limiting the key pool size to make it contain keys of only nodes in the cluster. Usage of a sensor node which has two types of radio communications helps the sensors to communicate with other cluster sensors through the cluster head which operates in the long range mode. The simulation performed confirmed the overall improved performance of the system.